# Democratizing Cross-Project Knowledge for Transforming Water Governance and Climate Resilience: an Agentic AI Ecosystem

Oriol Alàs[1*][0009-0001-6875-922X], Ian Palacín[1*][0009-0000-0148-7437], Iván Cester[1][0009-0005-6326-0146] and Edgar Rubión[1][0000-0003-3886-3170]

[1]Eurecat, Centre Tecnològic de Catalunya, Barcelona, Spain
edgar.rubion@eurecat.org

**Abstract.** EU-funded projects produce extensive evidence in the form of deliverables, reports, and case studies. However, this knowledge is typically scattered across projects and document structures, making it difficult to locate relevant, contextual information when making decisions. This paper introduces an Agentic AI Ecosystem that consolidates heterogeneous project documentation into a domain-aware knowledge base and provides adaptive access through an agentic Retrieval-Augmented Generation pipeline. Documents from three EU-funded projects (IMPETUS, NATALIE, and InnWater) are reconstructed into hierarchical section trees and indexed with structured metadata. Retrieval is supported by two domain-specific embedding models fine-tuned on domain data (water and climate). On top of this knowledge base, a ReAct-style controller iteratively orchestrates retrieval actions by allocating retrieval budget across domain collections and accumulating evidence until a stopping criterion is met. In a retrieval benchmark over the corpus, the proposed approach improves Mean Reciprocal Rank by 26.7% (relative) compared to a single-domain embedding baseline, indicating more consistent retrieval of relevant evidence to support cross-project knowledge synthesis and climate-resilient decision-making.

## 1 Introduction

Horizon Europe and related EU programmes have generated substantial technical evidence for climate adaptation and water governance in the form of technical deliverables and reports. In practice, these outputs are distributed across projects and embedded in long documents with heterogeneous structures, which makes it costly for local authorities, NGOs, and policymakers to identify the specific, contextual fragments needed for planning and decision-making. As recognized by the EU Climate Adaptation Strategy[1] and the IPCC Sixth Assessment Report[2], knowledge emerging from climate and environmental research must be not only scientifically robust but also accessible, contextualized, and inclusive to empower informed decision-making at all levels.

[*] Both authors contributed equally to this paper.
[1] European Commission, (2021), EU Strategy on Adaptation to Climate Change.

Retrieval-Augmented Generation (RAG) provides a natural paradigm for querying such corpora by grounding model outputs in external sources [1]. However, standard one-shot RAG pipelines remain brittle in cross-project settings. Relevant evidence may be spread across multiple documents and projects, section-level context is often required for correct interpretation, and user queries frequently span multiple domains (e.g., climate adaptation and water governance). Under these conditions, a single retrieval call may either miss key context or surface partially relevant fragments without mechanisms to iteratively refine the search or consolidate evidence across sources [2]. Recent work has addressed adjacent limitations in two directions. First, hierarchical and tree-based retrieval methods build multi-level representations to retrieve both local detail and global context from long documents, mitigating the shortcomings of flat chunking [3]. Second, agentic and adaptive RAG variants interleave reasoning with retrieval actions, deciding when to retrieve, refining queries or tool parameters, and accumulating evidence before answering, rather than committing to a fixed evidence set from a single call [4]. Despite this progress, cross-project decision-support scenarios impose an additional constraint: retrieval must be both domain-aware and provenance-preserving, with explicit control over how retrieval effort is allocated across heterogeneous project collections.

Generic embedding models often underperform on specialized project corpora where key distinctions depend on domain terminology; synthetic query-passage generation can provide supervision for retriever adaptation when labeled data are scarce [5]. To train the models efficiently under practical compute constraints, parameter-efficient fine-tuning methods such as Low-Rank Adaptation (LoRA) offer a common strategy to update a small number of trainable parameters while keeping the base model largely frozen [6].

To address these limitations, we propose an Agentic AI Ecosystem for cross-project knowledge access, demonstrated on documentation from three EU-funded projects: IMPETUS[3], NATALIE[4], and InnWater[5]. The ecosystem combines (i) a domain-aware knowledge base that reconstructs project documentation into hierarchical section trees and indexes fragments with structural and provenance metadata, and (ii) an agentic orchestration layer that performs adaptive, multi-step retrieval over domain-specific collections.

The main contributions of this work are: (1) a cross-project, domain-aware knowledge base that preserves document hierarchy and provenance metadata; (2) two domain-specific embedding spaces trained with evidence-grounded synthetic supervision, improving the discoverability of policy-relevant knowledge expressed in project-specific and technical language; (3) an iterative retrieval controller that allocates retrieval budget across domain collections via tool parameterization and evidence accumulation, supporting cross-sectoral and multi-project knowledge synthesis; and (4) an empirical evaluation over the projects corpus showing a 26.7% relative improvement in Mean Reciprocal Rank compared to a single-domain embedding baseline, indicating more reliable access to relevant evidence for decision-support scenarios.

---

[2] AR6 Synthesis Report (SYR), (2023)
[3] IMPETUS, Grant No. 101037084, https://climate-impetus.eu
[4] NATALIE, Grant No. 101112859, https://www.natalieproject.eu
[5] InnWater, Grant No. 101086512, https://www.innwater.eu

## 2 System Architecture

The proposed Agentic AI Ecosystem is designed to enable contextual, cross-project knowledge access across heterogeneous climate and water governance documentation. The architecture comprises two tightly coupled components: (i) a domain-aware knowledge database, responsible for ingesting, structuring, and indexing project documentation and (ii) a ReAct-style agent orchestration layer, which dynamically coordinates retrieval actions across domain-specific collections.

### 2.1 Domain-aware Knowledge Database

The ingestion pipeline turns unstructured and heterogeneous project documentation into a structured, section-aware knowledge base that preserves the original logical organization of each document. Our approach explicitly reconstructs the document's hierarchical structure and uses this structure as the backbone for downstream retrieval. The content is then indexed in a vector database using two domain-specific embedding models fine-tuned using project materials.

**Multimodal structure detection and traditional content extraction.**

Document ingestion employs a dual-parsing approach that combines AI-driven structural analysis with robust content extraction. A multimodal large language model (Google Gemini) is used to infer document structure through multimodal analysis, identifying section hierarchies, heading levels and organizational patterns. This approach improves robustness when processing documents with inconsistent formatting, visual noise, or irregular layouts, where rule-based parsers typically fail. In parallel, OpenParse, an open-source library designed for document parsing, is used to extract raw textual content, tables, and visual elements while preserving layout information. This enables filtering of non-content artefacts such as headers or page numbers and ensures accurate content recovery. Outputs from both parsers are aligned and merged into a unified representation by matching structurally inferred sections with their corresponding extracted content.

**Hierarchical Content Representation and Summarization**

The merged output is organized into a content-tree, where each node corresponds to a document section or subsection, and contains both raw content and associated metadata (project, document identifier and section path). To support retrieval at multiple levels of abstraction, the system generates hierarchical summaries using a depth-first search (DFS) post-order traversal. Leaf nodes are summarized directly from their raw text, while parent nodes recursively aggregate their own content with the summaries of their children.

**Domain Specific Embedding**

Generic pre-trained embedding models often fail to capture domain-specific terminology and relationships crucial for specialized retrieval tasks [7]. To address this limitation, we fine-tune two domain-specific embedding models (one for water governance and one for climate adaptation) that better align with the semantic spaces.

*Silver Dataset Construction*

To obtain supervision without manual labeling, a synthetic data generation pipeline was implemented to create high-quality question-answer-passage (QAP) triplets automatically. This pipeline utilizes Google Gemini to ingest raw project documentation and generate QAP samples only when they can be validated by verbatim evidence present in the text. Each generated entry consists of a) a concise question, b) a direct answer derived exclusively from the document content and c) extracted passage, consisting of exact quotes from the source text that serve as ground truth. Few-shot technique was applied to improve quality and style of the responses of the model. This process yielded 1064 high-quality triplets for the water domain and 846 for the climate domain.

*Fine-tuning*

To adapt the vector space to specific semantics of the water and climate domains, the multilingual BGE-M3 embedding model was fine-tuned (per domain) due to its multilingual coverage. Fine-tuning relied on a contrastive learning objective designed to improve semantic alignment for relevant query–passage pairs (positives) while decreasing similarity with non-relevant passages (negatives). For each query $i$, negatives were selected from a candidate set $\mathcal{N}(i)$. Let $q_i$ denote the embedding of query $i$, $p_i$ the embedding of its positive passage, and $\mathcal{N}(i)$ the set of negative passages considered for query $i$. The contrastive loss is defined as:

$$\mathcal{L} = -\frac{1}{N}\sum_{i=1}^{N} \log \frac{\exp\left(\frac{cos(q_i, p_i)}{\tau}\right)}{\sum_{d \in N(i)} \exp\left(\frac{cos(q_i, p_j)}{\tau}\right)} \quad (1)$$

where $N$ is the batch size, $\tau > 0$ is a temperature hyperparameter controlling the sharpness of the similarity distribution and $\cos(\cdot,\cdot)$ is defined as the cosine similarity function. Given the computational constraints and the risk of catastrophic forgetting associated with full-model fine-tuning, LoRA was adopted. This approach significantly reduces the number of the trainable parameters while retaining the linguistic capabilities of the base model.

**Two-Stage Retrieval**

Retrieval exploits hierarchical document structure through a two-stage procedure. Let $Q$ denote the user query and let each document be represented as a section tree, where each node has a summarized version of their children's sections and its own content.

*Document-level scoping*

The search space is reduced by computing semantic similarity between $Q$ and the root embedding of each document. Only documents whose scores exceeds a threshold $\theta_{doc}$ are retained.

*Node-level scoping*

For each shortlisted document, candidate nodes are scored using two complementary representations: (i) the embedding of the node's raw content and (ii) the node's generated summary. A node is selected as evidence if either similarity score exceeds a node threshold $\theta_{node}$, subject to a retrieval budget $k$. The output of retrieval is a set of evidence fragments $E$, where each fragment includes the text span plus hierarchical metadata to preserve provenance and support contextual synthesis. This retrieval operator is exposed to the agent as a parametrizable action:

$$\text{Retrieve}_d\big(Q;\ k_d, \theta_{doc,d}, \theta_{node,d}\big) \longrightarrow\ E_d \quad (2)$$

where $d$ denotes the target domain collection (e.g., water or climate) and $(k_d, \theta_{\cdot,d})$ control the retrieval budget and selectivity per domain.

## 2.2 Agent orchestration layer

The agent policy follows an iterative control loop that decides which retrieval action to execute, accumulates evidence across iterations, and stops to produce the final answer. Let $D$ denote the set of available domain collections. Given a query $Q$, the loop maintains an evidence set $E_t$ (initially empty). At each iteration $t < n$, the controller selects either (i) a retrieval action over one or more domain collections by choosing $D_t \subseteq D$ and invoking $\text{Retrieve}_d$ with domain-specific parameters (e.g., $k_d, \theta_{doc,d}, \theta_{node,d}$), or (ii) a STOP action to finalize the answer. Evidence is updated by accumulation:

$$E_t = E_{t-1} \cup \bigcup_{d\in D_t} Retrieve_d\big(Q; k_d, \theta_{doc,d}, \theta_{node,d}\big) \quad (3)$$

where each retrieved fragment includes text and structured metadata. The process terminates upon STOP or when the iteration budget $n$ is reached, after which the answer $A$ is generated conditioned on $(Q, E_t)$ with references to supporting fragments.

## 3 Experimental Evaluation

This section evaluates the impact of domain-specific embedding fine-tuning by comparing a generic embedding model within an information retrieval setup. The evaluation focuses on the distinction of the embedding space topology and their effect on downstream retrieval performance.

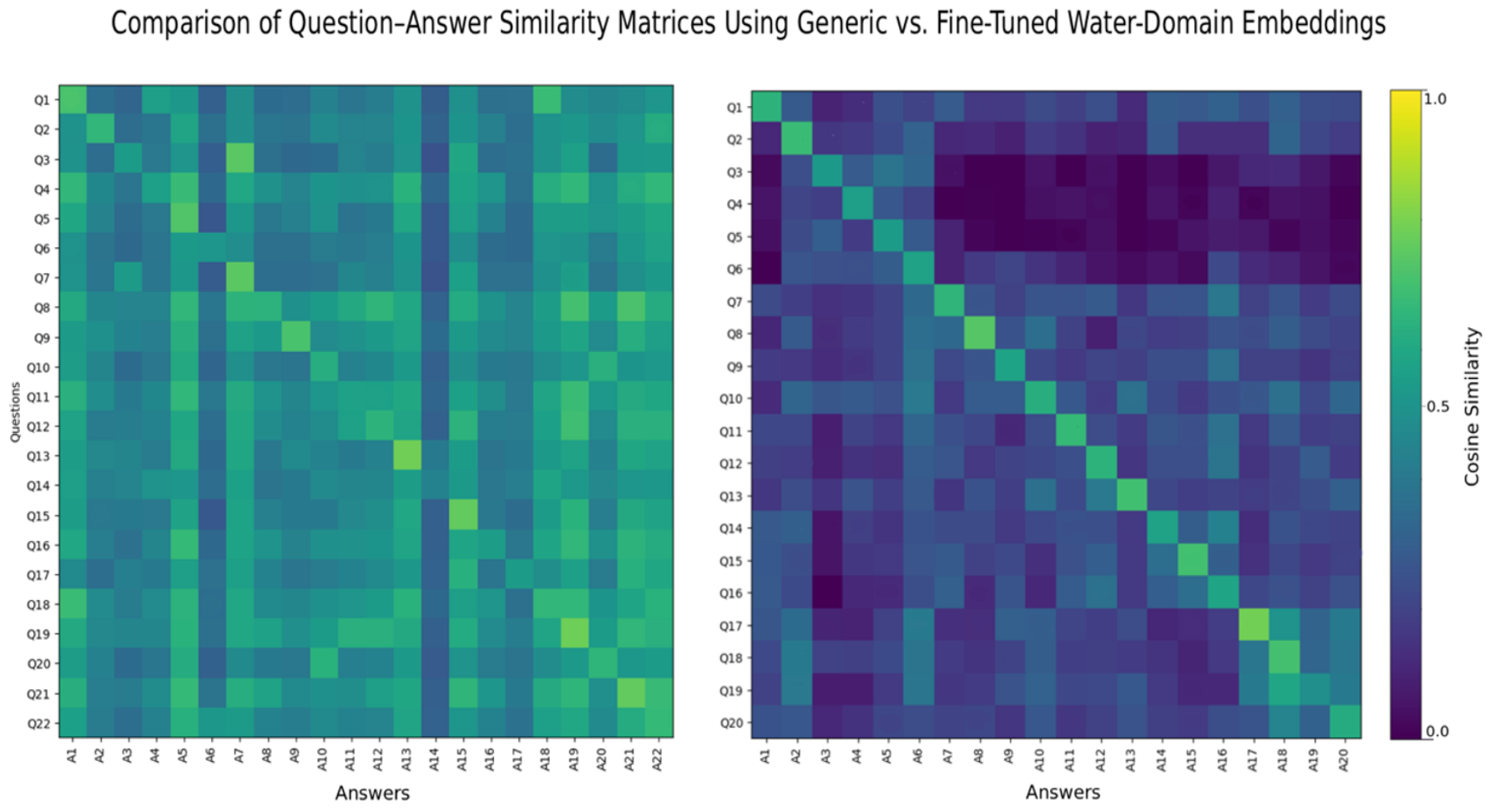

**Fig. 1.** Cosine Similarity Matrices for Question-Answer pairs from the water-domain validation set. Each row $i$ represents a synthetic question, and each column $j$ represents a corresponding answer fragment.

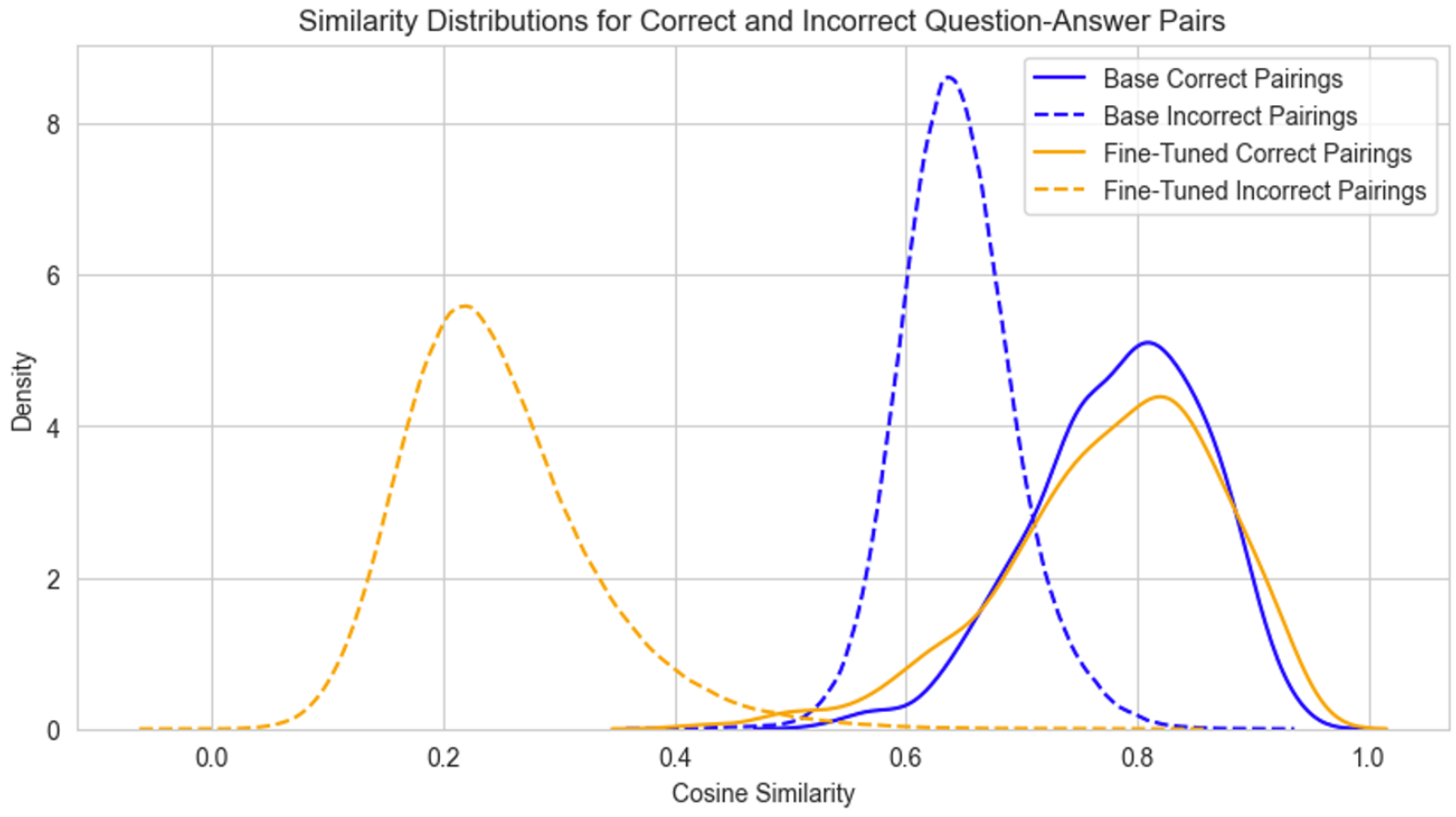

**Fig. 2.** Kernel density of cosine similarity for correct and incorrect question–answer pairings using the base (generic) and fine-tuned water-domain embedding models.

To evaluate the impact of fine-tuning, we conducted a controlled similarity analysis using a held-out validation set consisting of QAP triplets. Figure 1 illustrates the resulting cosine similarity matrices, where each row represents a specific query and each column corresponds to a potential answer fragment. The diagonal elements ($i = j$) denote correct pairs, while off-diagonal elements ($i \neq j$) represent incorrect matches. In this visualization, a perfect retrieval model would concentrate all similarity on the diagonal and zero similarity elsewhere. The matrix for the generic model shows a diffusion distribution with low contrast and significant noise; the similarity values for correct pairs (the diagonal) are often indistinguishable from those of incorrect pairs (the off-diagonal noise). Following domain-specific fine-tuning, the embedding space undergoes a clearer separation: the diagonal becomes more prominent and the off-diagonal "background noise" is globally suppressed. This indicates that the fine-tuned model has learned to better distinguish relevant from irrelevant evidence, effectively increasing the "margin" between correct and incorrect retrievals. The structural shift is further quantified in Figure 2. In the generic model, the similarity distributions for correct and incorrect pairs show a high degree of overlap, which leads to retrieval ambiguity. In contrast, fine-tuning effectively decouples these distributions: it maintains high similarity for correct matches while shifting the “noise” distribution toward lower values. This dual evidence demonstrates that the model has successfully learned to isolate relevant evidence from irrelevant information.

Finally, Mean Reciprocal Rank (MRR) [8] was used to quantify retrieval effectiveness. For each query, the reciprocal rank is defined as the inverse of the rank position at which the first correct answer appears in the retrieved list. Formally, for a set of $N$ queries, MRR is computed as (4):

$$MRR = \frac{1}{N}\sum_{i=1}^{N}\frac{1}{rank_i} \tag{4}$$

where $rank_i$ denotes the position of the first relevant answer for the $i$-th query. MRR places greater emphasis on retrieving the correct answer at higher ranks, assigning the maximum score when the correct item appears at the top position and lower score as the rank increases. The fine-tuned embedding model achieves a 26.7% relative improvement compared with the baseline RAG system. This improvement reflects both an increased frequency of correct answers appearing in top-ranked positions and a reduction in average rank across the evaluation set. In practical terms, relevant contextual fragments are retrieved earlier, providing information retrieval tools with more domain-aligned evidence.

## 4 Conclusion

By transforming static project documentation into a living knowledge ecosystem that evolves as new projects emerge, this work helps address the urgent need identified in EU policy frameworks to make research outputs accessible and actionable for local decision-makers, ultimately supporting the transition toward climate-resilient societies. Experimental results demonstrate a 26.7% relative improvement in MRR over a single-domain baseline, confirming that domain alignment and hierarchical scoping materially enhance early retrieval precision and discriminative separation in the embedding space. Building upon these findings, future research will focus on expanding the ecosystem's capabilities through three primary avenues. First, the system will be extended with multi-modal retrieval capabilities to support richer evidence grounding and multimodal retrieval. Second, the ecosystem will be extended to additional EU-funded projects and adjacent policy domains to assess scalability, cross-domain generalization and embedding transferability. Finally, human-in-the-loop evaluations are planned to assess the faithfulness and reasoning quality of the agentic responses in the real-word policy-making scenarios.

**Acknowledgements**

This work has received funding from the European Union's Horizon 2020 programme under grant agreement No. 101037084 (IMPETUS) and from the European Union's Horizon Europe research and innovation programme under grant agreements No. 101112859 (NATALIE) and No. 101086512 (InnWater). The views and opinions expressed are those of the authors only and do not necessarily reflect those of the European Union or the European Commission. Neither the European Union nor the granting authority can be held responsible for them.